\documentclass[10pt,journal,final,finalsubmission,compsoc]{IEEEtran}

\usepackage{t1enc}
\usepackage[utf8]{inputenc}
\usepackage{graphicx}
\usepackage{epstopdf}
\usepackage{color}
\usepackage[normalem]{ulem}

\usepackage[nocompress]{cite}
\usepackage{soul}

\begin{document}

\title{``Pull moves'' for rectangular lattice polymer models are
not fully reversible}

\author{Dániel Györffy, Péter Závodszky and András Szilágyi}

\IEEEcompsoctitleabstractindextext{%
\begin{abstract}
``Pull moves'' is a popular move set for
lattice polymer model simulations. We show that the proof given for
its reversibility earlier is flawed, and some moves are irreversible,
which leads to biases in the parameters estimated from the
simulations. We show how to make the move set fully reversible.
\end{abstract}

}

\pagestyle{empty}

\onecolumn

\vspace*{5cm}

\begin{center}
\Huge
``Pull moves'' for rectangular lattice polymer models are
not fully reversible
\end{center}

\linespread{1.2}

\Large

\begin{center}
    Dániel Györffy, Péter Závodszky and András Szilágyi
\end{center}

\vspace{2cm}

\noindent \copyright 2012 IEEE. Personal use of this material is
permitted. Permission from IEEE must be obtained for all other uses,
in any current or future media, including reprinting/republishing this
material for advertising or promotional purposes, creating new
collective works, for resale or redistribution to servers or lists,
orreuse of any copyrighted component of this work in other works.

\vspace{2cm}

\noindent This paper has been accepted for publication in
\emph{IEEE/ACM Transactions on Computational Biology and
Bioinformatics}. The manuscript was first submitted on March 29, 2012,
revised August 5, 2012, accepted September 29, 2012.

\twocolumn

\linespread{1.}
\normalsize

\pagestyle{plain}

\pagenumbering{arabic}

\maketitle

\IEEEdisplaynotcompsoctitleabstractindextext

\section{Introduction}

Lattice polymer models have long been used for theoretical studies of
various aspects of polymer behavior \cite{1525, 1527}. In the HP
lattice model of proteins \cite{695}, a chain is represented
as a self-avoiding walk on a rectangular (square or cubic) lattice and
the sequence space is reduced to two types of beads: a hydrophilic one
(denoted as P) and a hydrophobic one (denoted as H). Analyses carried
out on HP lattice protein models provided the foundations of the
modern theory of protein folding \cite{172}.

For short polymers, exhaustive enumeration can be carried out, but for
longer chains, excessive computational requirements permit only
sampling the conformational space. Several efficient sampling
methods have been developed such as the Metropolis--Hastings Monte
Carlo method \cite{1196}, equi-energy sampling \cite{547} and
Wang--Landau sampling \cite{1203, 1200}. The main goal of the
sampling may be finding the conformation with the lowest energy,
generating a canonical ensemble of conformations, or estimating
various parameters such as the density of states, free energies,
thermodynamic averages, etc.

\begin{figure}
    \centering \includegraphics{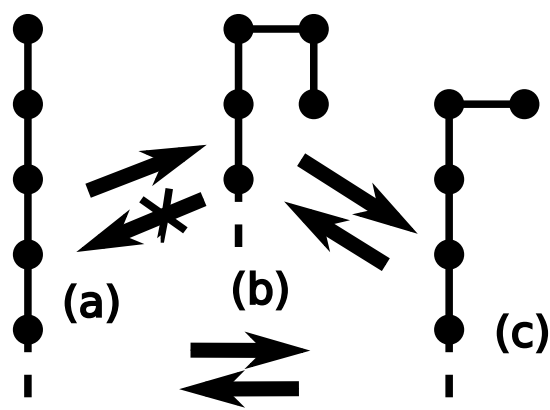} 
    \caption{\label{irreversibility}An
      example for an irreversible pull move. Conformation (a) is a chain
      with a straight end. A particular end move results in a hook at the
      end of the chain, conformation (b). But a reverse move is not possible
      because pulling the chain in the other direction results in a chain
      with a bent end (c) instead of a straight end. Excluding the move from
      (a) to (b) will not affect ergodicity because it remains possible to
      get from (a) to (b) by two moves, through (c). Note that the figure
      only shows an example; all end moves that result in a hooked end
      are irreversible.}
\end{figure}

All of these sampling methods require a mutation step. For rectangular
lattice models, several move sets have been developed to serve as a mutation
operator. In pivot moves \cite{1529,1475}, one of the beads serves as a
pivot point and a symmetry operation is carried out. Pivot moves are ergodic
but due to the significant change in conformation, the acceptance rate is
very low for compact conformations \cite{1475,1462}. In the class of {\em
k}-bead moves, a predefined number ({\em k}) of contiguous beads are
relocated \cite{1527}, but these moves are not ergodic for longer chains
\cite{1474}. A fundamentally different method is bond-rebridging
\cite{1458}, which provides a high acceptance rate at the cost of not being
ergodic \cite{1531, 1462}.

Lesh and coworkers introduced a new local move set, called pull moves
\cite{1205}, and provided proof that the move set is ergodic and reversible.
There are two types of pull moves. In one type, a four-bead loop is formed
and the chain is pulled until the end is reached or an existing loop is
straightened. To ensure reversibility, another move type was introduced
where one end of the chain is pulled until a loop is eliminated \cite{1205,
547}. ``Pull moves'' has become a popular and commonly used move set for
lattice polymer simulations. It has been used in replica
exchange Monte Carlo simulations \cite{1199} and applications of
equi-energy \cite{547} and Wang--Landau sampling \cite{1200, 1531, 1456,
1462} and tabu search \cite{1205,1605}. It has been adapted for triangular
and hexagonal lattice polymers \cite{1545, 1463}.

\section{Results}

Applications of pull moves have been implemented assuming that the move set
is reversible (i.e. that the reverse of each valid move from state A to
state B is also a valid move from B to A), based on the proof of
reversibility given by Lesh et al. \cite{1205}. However, examining the
proof, we noticed that it contains a flaw as it overlooks the fact that one
type of end move is actually not reversible. (The irreversibility does not
affect pull moves on triangular and hexagonal lattices.)

Here, we provide a proof that one type of end pull moves is irreversible.
Let us number the beads of the chain from $1$ to $n$. The irreversibility
affects those end moves that result in a hook at the end of the chain as
shown in Figure \ref{irreversibility}(b). By this move, both beads $1$ and
$2$ move to a new location such that bead $1$ will be adjacent to the old
location of bead $2$. According to the rules of pull moves described in Lesh
et al. \cite{1205}, this move can be reversed by pulling the chain in the
other direction until a valid conformation is reached. However, when we try
to perform this reverse move, bead $2$ will return to its original position,
but because this position is adjacent to the new location of bead $1$, the
chain will be in a valid conformation without moving bead $1$. Thus, bead
$1$ is not returned to its original position. The same reasoning applies to
beads $n$ and $n-1$. Figure \ref{irreversibility} shows one example of the
irreversibility where, starting from a chain with a straight end, one move
results in a hook at the end of the chain, but the supposed reverse move
produces a chain with a bent end.

For a 2D chain with length $L$, the maximum number of possible
internal and end pull moves is $4(L-2)$ and $18$, respectively; for 3D
chains, the corresponding numbers are $8(L-2)$ and $25$. Four of the $18$
end pull moves in 2D, and $8$ of the $25$ end pull moves in 3D are
irreversible. We calculated that $15\ \%$ of edges in the transition graph
of a 2D 10-bead chain represent an irreversible transition.

Monte Carlo sampling schemes, including Wang--Landau
sampling, rely on the principle of detailed balance which requires
microscopic reversibility. When microscopic reversibility is violated,
cycles of microstates will arise, and statistical weights derived from the
simulations will be incorrect.

Fortunately, the irreversibility can easily be fixed by simply
excluding the irreversible moves, i.e. those moves that produce a hook
at the end of the chain like that depicted in Figure
\ref{irreversibility}. This new, reduced move set preserves
ergodicity; this is easily seen in Figure \ref{irreversibility} as
direct moves from (a) to (c) and (c) to (a) are both allowed.

If the sampling algorithms are implemented without simplifications,
the irreversible moves will never be accepted, leading only to a waste
of CPU time. However, implementations often use the assumption of
reversibility to introduce simplifications, leading to irreversible
moves being accepted, and thus, a bias in the sampling.

In Monte Carlo samplings, such as e.g. Wang--Landau
sampling, the transition probability is proportional to the rates of
the {\em a priori} probabilities of the transition from one state to
the other and {\em vice versa}:
\begin{equation}
    \label{transition}
    p \left( A \rightarrow B \right) \propto \frac{p_{ap}
    \left( B \rightarrow A \right)}{p_{ap} \left( A \rightarrow B
    \right)}.
\end{equation}
The {\em a priori} probability of a transition is
\begin{equation}
    p_{ap} \left( A \rightarrow B \right) = \frac{n \left( A
  \rightarrow B \right)}{n \left( A \right)}
\end{equation} 
where $n \left( A \rightarrow B \right)$ is the number of transitions from
the state $A$ to state $B$ and $n \left( A \right)$ is the number of
transitions originating from $A$.  If a move is irreversible such that $n(B
\rightarrow A)=0$ (as is the case for the irreversible end moves), then $p(A
\rightarrow B)$ will be zero and this move will never be accepted during the
simulation. Thus, CPU time will be wasted but the sampling will be correct.

However, when implementing the sampling algorithm, a simplified expression
for $p(A \rightarrow B)$ may be used instead of Equation (\ref{transition}).
If the move set is assumed reversible then
\begin{equation} 
    \label{reversibility}
    n \left( A \rightarrow B \right) = n \left( B \rightarrow A \right),
\end{equation} 
and we can simplify Equation (\ref{transition}) so that 
\begin{equation} 
    \label{simplification}
    p \left( A \rightarrow B \right) \propto \frac{n \left( A
    \right)}{n \left( B \right)} . 
\end{equation} 
With an irreversible move set, Equation
(\ref{reversibility}) is not valid and Equation (\ref{simplification})
yields nonzero probabilities even for irreversible moves. This leads to
accepting irreversible moves and, consequently, incorrect sampling.

\begin{figure}[h!]
    \centering \includegraphics{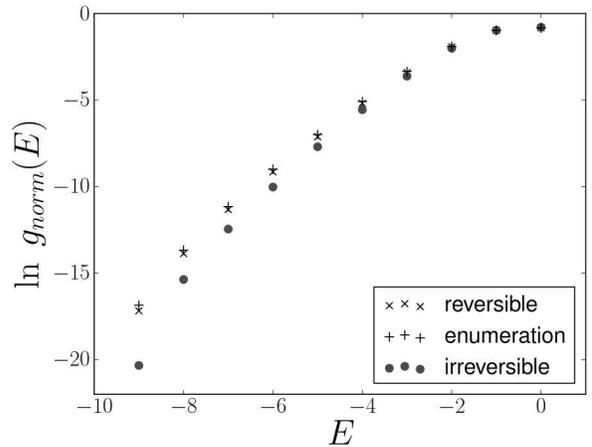}
    \caption{\label{comparison}Logarithm of the normalized density of states
    ($g_{norm} \left( E \right)$) for the 2D 20mer HPHPPHHPHPPHPHHPPHPH,
    also investigated in \cite{1472}, from Wang--Landau simulations (50 runs
    averaged) with reversible and irreversible pull moves, and from
    exhaustive enumeration. Wang--Landau sampling was stopped when
    the logarithm of the modification factor fell below $10^{-8}$. The
    flatness criterion was chosen to be $p=0.8$} 
\end{figure}

To investigate the effect of neglecting the irreversibility, we estimated
the density of states of a 2D 20mer by Wang--Landau simulations
\cite{1203} using both the original (irreversible) and the fixed
(reversible) move set (applying the simplification in Equation
(\ref{simplification})), and compared them with the exact values obtained by
exhaustive enumeration (Figure \ref{comparison}). As the figure shows, by
using the irreversible moves, the density of states is significantly
underestimated for most energy levels, especially for the lowest ones. We
obtained similar but less marked differences for a 2D 64mer, a 3D 103mer,
and a 18mer homopolymer also investigated previously \cite{1472, 1456, 1457}.

\section{Discussion}

We have shown that some end pull moves are irreversible, and ignoring the
irreversibility may lead to inaccurate sampling. Whether and to what extent
this irreversibility affected the results of earlier studies is an open
question. Although we cannot rule out the possibility that some authors
recognized the irreversibility and excluded the irreversible moves, all
papers we examined state that pull moves are reversible and we found no
mention of excluding any moves. Even if irreversible moves were not
explicitly excluded in a study, this may not have had a noticeable effect on
the results for several reasons: (i) the study only used pull moves for
finding the ground state, so the irreversibility did not affect it
\cite{1205, 1199, 1605}; (ii) the authors may have used the full formula
(Equation (\ref{transition})) to calculate transition probabilities, thus
the irreversible moves were tried but always rejected; (iii) the study was
done on long chains where end moves represent only a small fraction of all
moves, thus their irreversibility only had a small effect.

In many cases, studies do not provide enough details on the implementation
of the used algorithm and their results to allow us an assessment of whether
and how irreversibility may have affected the results. A few studies
\cite{547, 1200, 1457} provide an accuracy analysis, comparing the density
of states obtained from the simulations with that from exact enumeration
\cite{547, 1200}, and they only find small differences from the exact
values. However, they do not state whether they used the full formula
(Equation (\ref{transition})) or its simplified form (Equation
(\ref{simplification})); only the simplified form leads to inaccuracies. On
the other hand, studies that explicitly state that they used the simplified
formula (Equation (\ref{simplification})) \cite{1456, 1531} do not provide
an accuracy analysis (e.g. because of working with long chains where
enumeration is not possible), so it is impossible to judge the accuracy of
their results. An exception is the study of Swetnam and Allen \cite{1457},
which both calculates accuracy and states that the simplified formula
(Equation (\ref{simplification})) was used. However, their calculations
are for a 18mer homopolymer where, according to our own calculations, the
difference caused by the irreversibility is small. Besides, they use various
algorithms with varying results, and we are unable to directly compare their
results with ours. Thus, we cannot determine with certainty whether
potential irreversibility affected their results.

There is a publicly available software package, LatPack \cite{1563}, that
implements pull moves. We tested the program and found that it fails to
exclude the irreversible end pull moves.

\section{Conclusion}

As our demonstrations indicate, accurate calculations in the future
should be based on simulations using the fixed version of pull moves
as described in the present article. The simple recipe is that all end pull
moves that result in a hook at the end of the chain should be excluded.


\end{document}